\documentclass[twocolumn]{article}  
\usepackage{graphicx}
\usepackage[varg]{txfonts}

\newcommand{\hcn}{\ensuremath{\mathrm{HCN}}}
\newcommand{\hnc}{\ensuremath{\mathrm{HNC}}}
\newcommand{\hthicn}{\ensuremath{\mathrm{H^{13}CN}}}
\newcommand{\hnthic}{\ensuremath{\mathrm{HN^{13}C}}}
\newcommand{\hfifnc}{\ensuremath{\mathrm{H^{15}NC}}}
\newcommand{\hcfifn}{\ensuremath{\mathrm{HC^{15}N}}}
\newcommand{\diaz}{\ensuremath{\mathrm{N_2H^+}}}
\newcommand{\ammo}{\ensuremath{\mathrm{NH_3}}}
\newcommand{\cyana}{\ensuremath{\mathrm{HC_3N}}}
\newcommand{\molh}{\ensuremath{\mathrm{H_2}}}
\newcommand{\ceio}{\ensuremath{\mathrm{C^{18}O}}}

\newcommand{\kmps}{\ensuremath{\mathrm{km\,s^{-1}}}}
\newcommand{\Msun}{\ensuremath{\mathrm{M}_\odot}}

\newcommand{\percmsq}{\ensuremath{\mathrm{cm^{-2}}}}

\newcommand{\hour}{\ensuremath{^\mathrm{h}}}
\newcommand{\minute}{\ensuremath{^\mathrm{m}}}
\newcommand{\second}{\ensuremath{^\mathrm{s}}}

\begin{document}

\title{HCN and HNC mapping of the protostellar core Cha-MMS1
\thanks{Based on observations collected at the European Southern
Observatory, La Silla, Chile, and at the Parkes Observatory operated
by the Australia Telescope National Facility (ATNF).}}

   \author{P.P. Tennekes$^{1,2}$  \and
          J. Harju$^{1}$
          \and M. Juvela$^{1}$ \and
          L.V. T\'{o}th$^{3,4}$}

\maketitle

  $^1$Observatory, P.O. Box 14, 
              FI-00014 University of Helsinki, Finland    

  $^2$Julius Institute, Utrecht University, 
          The Netherlands
               
  $^3$Max-Planck-Institut f\"ur Astronomie (MPIA), K\"onigstuhl 17, 
       69117 Heidelberg, Germany

  $^4$Department of Astronomy of the Lor{\'a}nd E{\"o}tv{\"o}s University, 
      P{\'a}zm{\'a}ny P{\'e}ter s{\'e}t{\'a}ny 1, 1117 Budapest, Hungary

  \begin{abstract} 

  {\sl Aims.}
{The purpose of this study is to investigate the distributions of the
isomeric molecules HCN and HNC and estimate their abundance ratio in
the protostellar core Cha-MMS1 located in Chamaeleon {\sc i}.}

  {\sl Methods.}
{The core was mapped in the $J=1-0$ rotational lines of \hcn, \hnc,
and \hnthic. The column densities of \hthicn, \hnthic, \hfifnc\
and \ammo\ were estimated towards the centre of the core.}

  {\sl Results.}
{The core is well delineated in all three maps. The kinetic
temperature in the core, derived from the \ammo\ (1,1) and (2,2)
inversion lines, is $12.1 \pm 0.1$ K.  The \hnthic/\hthicn\ column
density ratio is between 3 and 4, i.e. similar to values found in
several other cold cores.  The \hnthic/\hfifnc\ column density ratio
is $\sim 7$.  In case no $^{15}$N fractionation occurs in \hnc\ (as
suggested by recent modelling results), the \hnc/\hnthic\ abundance
ratio is in the range $30-40$, which indicates a high degree of
$^{13}$C fractionation in \hnc.  Assuming no differential $^{13}$C
fractionation the \hcn\ and \hnc\ abundances are
estimated to be $\sim 7 \, 10^{-10}$ and $\sim 2\, 10^{-9}$,
respectively, the former being nearly two orders of magnitude smaller
than that of \ammo.  Using also previously determined column densities
in Cha-MMS1, we can put the most commonly observed nitrogenous
molecules in the following order according to their fractional
abundances: $\chi(\ammo) > \chi(\cyana) > \chi(\hnc) > \chi(\hcn) >
\chi(\diaz)$.} 

  {\sl Conclusions.}
{The relationships between molecular abundances suggest 
that Cha-MMS1 represents an evolved chemical
stage, experiencing at present the 'late-time' cyanopolyyne peak.  
The possibility that the relatively high HNC/HCN ratio derived
here is only valid for the $^{13}$C isotopic substitutes cannot be excluded
on the basis of the present and other available data.}

\end{abstract}

\section{Introduction}

Hydrogen cyanide, \hcn, and its metastable isomer hydrogen isocyanide,
\hnc, are commonly used tracers of dense gas in molecular clouds. The
\hnc\ column density has been observed to be similar or 
larger than that of \hcn\ in cold regions, whereas in warm GMC cores
the [HNC]/[HCN] ratio is typically much smaller than unity
(e.g., \cite{goldsmith1981}; \cite{churchwell1984}; \cite{irvine1984}; 
\cite{schilke1992}; \cite{hirota1998}). 

Gas-phase chemistry models, including both ion-molecule and
neutral-neutral reactions, predict that the \hnc/\hcn\ abundance ratio
should be close to unity in cold gas (see e.g. \cite{herbst2000},
where also previous work is reviewed).  The two isomers are thought to
form primarily via the dissociative recombination of HCNH$^+$, which
in turn is produced by a reaction between \ammo\ and C$^+$. The
recombination reaction regulates the abundance ratio despite the
initial production mechanisms of the two isomers, provided that they
are efficiently protonated via reactions with ions, such as H$_3^+$
and HCO$^+$ (\cite{churchwell1984}; \cite{brown1989}). Recent theoretical
studies corroborate the equal branching ratios for \hcn\ and \hnc\
in dissociative recombination (\cite{hickman2005}; \cite{ishii2006}).

\hnc/\hcn\ column density ratios clearly larger than unity have been
reported: \cite{churchwell1984} find that [\hnc]/[\hcn] may be as high
as $\sim 10$ in some dark clouds cores, and \cite{hirota1998} find a
ratio of $\sim 5$ in L1498.  The first result quoted is uncertain
because it depends on the assumed degree of $^{13}$C fractionation in
HNC, whereas the second value refers to \hnthic\ and \hthicn\
measurements. Nevertheless, both results suggest that the \hcn/\hnc\
chemistry is not fully understood.  Further determinations
of the relative abundances of these molecules towards cores
with well-defined chemical and physical characteristics seem therefore
warranted. 

In this paper we report on observations of \hcn\ and \hnc\ and some of
their isotopologues towards the dense core Cha-MMS1, which is
situated near the reflection nebula Ced 110 in the central part of the
Chamaeleon I dark cloud.  We also derive the gas kinetic temperature
in the core by using the (1,1) and (2,2) inversion lines of
\ammo. Cha-MMS1 was discovered in the 1.3 mm dust continuum by
\cite{reipurth1996}, and was studied in several molecules by
\cite{kontinen2000}. The object is embodied in one of
the most massive 'clumps' in Chamaeleon {\sc i} identified in the
large scale \ceio\ survey of \cite{haikala2005} (clump no. 3, $\sim 12
\Msun$). \cite{reipurth1996} suggested that Cha-MMS1 contains a
Class 0 protostar. A FIR source (Ced 110 IRS10) was detected near the
centre of the core by \cite{lehtinen2001}. However, no centimetre
continuum nor near-infrared sources have been found in its
neighbourhood (\cite{lehtinen2003} and references therein), and the
core therefore represents a very early stage of star formation.
At the same time, its chemical composition has probably reached an 
advanced stage (\cite{kontinen2000}). 

In Sect. 2 of this paper we describe the observations. In Sect. 3 we
present the direct observational results in the form of spectra and
maps, and derive the \hthicn, \hnthic, and \hfifnc\ column
densities. By combining our results with previous observations we
derive in Sect.~4 the fractional abundances of several nitrogen
containing molecules in Cha-MMS1, and discuss briefly the chemical
state of the core. In Sect. 5 we summarize our results.

\section{Observations}

The $J=1-0$ transitions of \hcn, \hthicn, \hnc, \hnthic\ and \hfifnc\
at about 90 GHz were observed with the
Swedish-ESO-Submillimeter-Telescope, SEST, located on La Silla in
Chile.  The observations took place in December 1990.  A 3~mm dual
polarization Schottky receiver was used in the frequency switching
mode.  The backend was a 2000 channel acousto optical spectrometer
with an 86\,MHz bandwidth. The velocity resolution with this
configuration is about $0.15\kmps$ at 90 GHz. Typical values for the
single sideband system temperatures ranged from 350~K to 450~K.  The
half-power beam width of the SEST is about $55^{\prime\prime}$ at the
frequencies observed.  Calibration was achieved by the chopper wheel
method. Pointing was checked every 2-3 hours towards a nearby SiO
maser source. We estimate the pointing accuracy to have been better
than $5^{\prime\prime}$ during the observations.  The focusing was done using
a strong SiO maser.

The (1,1) and (2,2) inversion lines of \ammo\ at about 23 GHz were
observed with the Parkes 64-m radio telescope of the Australian
Telescope National Facility (ATNF) in April 2000.  The FWHM of the telescope 
is about $80^{\prime\prime}$ at this frequency. The two lines were
observed simultaneously in two orthogonal polarizations in the
frequency switching mode, using the 8192 channel autocorrelator. The
calibration was checked by observing standard extragalactic
calibration sources at different elevations, and by comparing the
ammonia line intensities towards some strong galactic sources also
visible from the Effelsberg 100-m telescope, taking the different beam
sizes into account.

The electric dipole moments of the molecules and the line
frequencies used in this paper are listed in
Table~\ref{table:lineparameters}. Except for the frequencies of the
\hnc\ and \hnthic\ hyperfine transitions adopted from
\cite{frerking1979}, the parameters are obtained from the Jet
Propulsion Laboratory molecular spectroscopy data base
(\verb1http://spec.jpl.nasa.gov1), where also the original references
are given.  The hyperfine components of \ammo\ are not
listed. The frequencies given represent the centre frequencies of 18
and 21 individual components of the (1,1) and (2,2) lines,
respectively, distributed over a range of about 4 MHz.

\begin{table}
\begin{tabular}{lcccc}
molecule &  $\mu\,^a$ & \multicolumn{2}{c}{transition} & $\nu\,^a$    \\
         & (Debye)&         &                      & (MHz)    \\\hline
\hcn     & 2.98   &         & $F=1-1$              & 88630.416\\
         &        & $J=1-0$ & $F=2-1$              & 88631.847\\
         &        &         & $F=0-1$              & 88633.936\\

\hthicn  & 2.98   &         & $F=1-1$              & 86338.767\\
         &        & $J=1-0$ & $F=2-1$              & 86340.184\\
         &        &         & $F=0-1$              & 86342.274\\

\hnc$\,^b$ & 3.05  &         & $F=1-1$              & 90663.656\\
          &       & $J=1-0$ & $F=2-1$              & 90663.574\\
          &       &         & $F=0-1$              & 90663.450\\

\hnthic$\,^b$ &2.70&         & $F=1-1$              & 87090.942\\
             &    & $J=1-0$ & $F=2-1$              & 87090.859 \\
             &    &         & $F=0-1$              & 87090.735 \\

\hfifnc      &2.70& $J=1-0$ &                      & 88865.715\\

\ammo        &1.48 & $(J,K)=$   & $(1,1)$ & 23694.495 \\
             &     & $(J,K)=$   & $(2,2)$ & 23722.633 \\ \hline
\end{tabular}

$^a$ \verb1http://spec.jpl.nasa.gov1 \\
$^b$ Frequencies from \cite{frerking1979}

\caption{The permanent dipole moments of the molecules and the rest 
frequencies of the observed transitions.}
\label{table:lineparameters}
\end{table}

\section{Maps and column densities}
\subsection{Maps}

\begin{figure}
\includegraphics[height=7.4cm]{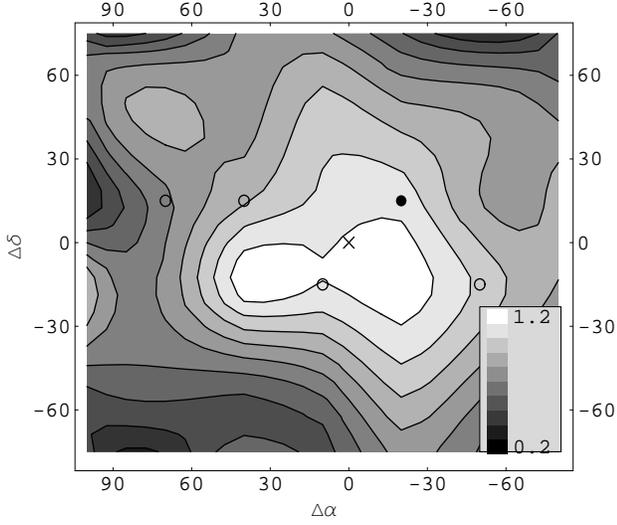}
\caption{Integrated \hcn($J=1-0$) line intensity map.  The integration
is performed over all three hyperfine structure components ($F=1-1$,
$F=2-1$ and $F=0-1$).  The map is centered on the position of
Cha-MMS1a (denoted by a cross), $\alpha_{2000} = 11 \hour 06 \minute
31.7\second$, $\delta_{2000} = -77 ^\circ 23^\prime
32^{\prime\prime}$, and the offsets are in arcseconds.  The contour
levels indicated with ten shades of grey and their values can be read
in the bottom right corner. The intensity unit is K km s$^{-1}$.  The
positions used for the column density estimates
($-20^{\prime\prime},20^{\prime\prime}$) and $T_{\rm ex}$ estimates
(see Table~\ref{table:texfromhcn}) are marked with a black dot and
open circles, respectively. The grid spacing used in the map is
$30^{\prime\prime}$.}
\label{figure:hcnmap}
\end{figure}

\begin{figure}
\includegraphics[height=7.4cm]{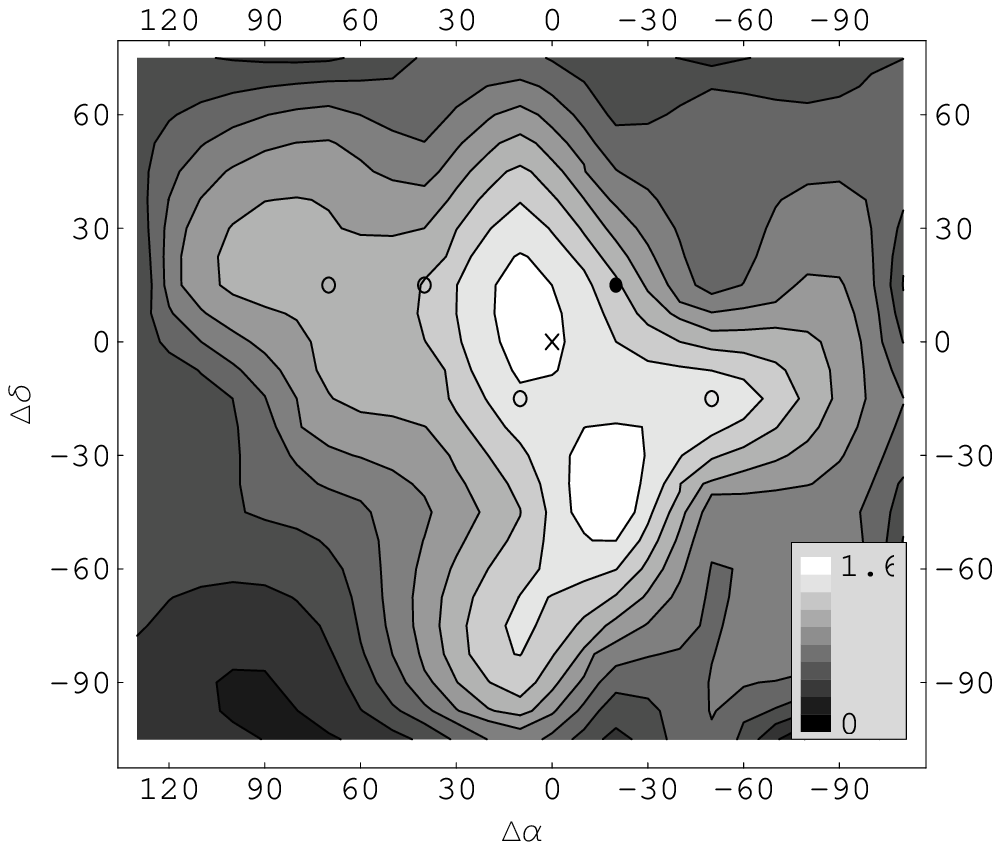}
\caption{Integrated \hnc($J=1-0$) line intensity map. The coordinates and
markers as in Fig~\ref{figure:hcnmap}.}
\label{figure:hncmap}
\end{figure}

\begin{figure}
\includegraphics[height=7.4cm]{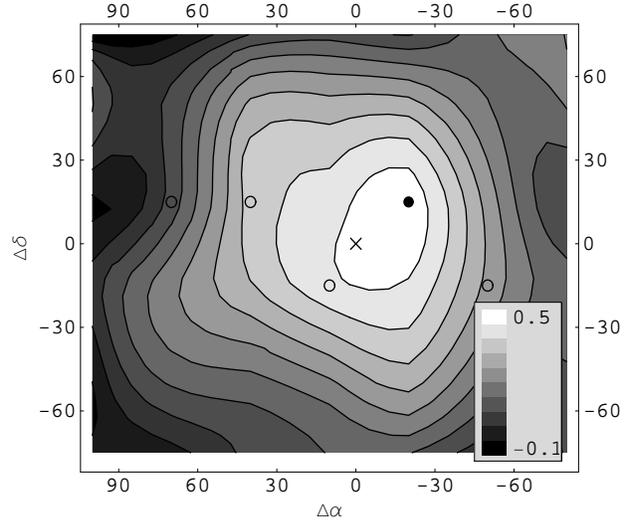}
\caption{Integrated \hnthic($J=1-0$) line intensity map. The coordinates and
markers as in Fig~\ref{figure:hcnmap}.}
\label{figure:hn13cmap}
\end{figure}

The integrated \hcn$(J=1-0)$, \hnc$(J=1-0)$ and \hnthic$(J=1-0)$
intensity maps of Cha-MMS1 are presented as contour plots in
Figs.~\ref{figure:hcnmap}, \ref{figure:hncmap} and
\ref{figure:hn13cmap}, respectively. The values of the contour levels
are given in the bottom right of each figure.  The position of the 1.3
mm dust emission maximum, Cha-MMS1a (\cite{reipurth1996}), is chosen
as the map centre. The coordinates of Cha-MMS1a are $\alpha_{2000} =
11 \hour 06 \minute 31.7\second$, $\delta_{2000} = -77 ^\circ
23^\prime 32^{\prime\prime}$.  The grid spacing used in the maps was
$30^{\prime\prime}$.  Also indicated in these figures are the location
of the \hnthic\ maximum ($\Delta\alpha,\Delta\delta =
-20^{\prime\prime},20^{\prime\prime}$, black dot), and the positions
used for the \hcn($J=1-0$) excitation temperature estimates in the LTE
method (open circles, see below).  Long integrations towards the
\hnthic\ maximum were made besides \hnc\ and \hcn\, also in the
$J=1-0$ lines of \hthicn\ and \hfifnc. The five spectra obtained
towards this position are shown in Fig.~\ref{figure:5spectra}.

The three maps have differences, but they all show a relatively
compact distribution peaking near the position of Cha-MMS1.  The
\hnthic\ emission with the lowest opacity reflects probably best the
distribution dense gas.

\subsection{LTE column densities} 

The column densities of the molecules have been determined with
two different methods. First, we have derived the optical thicknesses
of the lines and the total column densities of the molecules by
assuming local thermodynamic equilibrium, LTE. The LTE method involves
idealistic assumptions about the homogenity of the cloud.  In the next
subsection we present modelling of the observed spectra with a Monte Carlo
program using a realistic density distribution.

In the following, we assume that the rotational levels of all five
isotopologues are characterized by the same $T_{\rm ex}$, and that the core
is homogeneous, i.e.  the same $T_{\rm ex}$ is valid for different
locations. The former assumption can be justified by the similar
dipole moments and rotational constants of the molecules.  The second
assumption infers that the excitation conditions remain constant along
the line of sight and also between close-lying positions in the
core. 

\begin{figure}
\includegraphics[height=7.5cm,angle=270]{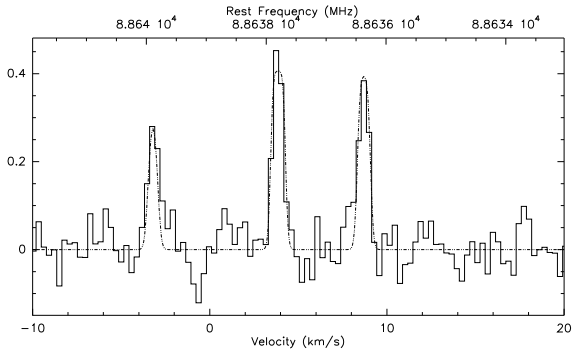}
\caption{The \hcn($J=1-0$) spectrum towards offset 
($40^{\prime\prime},20^{\prime\prime}$) used to determine the $T_{ex}$. 
The dotted line represents a fit to the hyperfine structure.}  
\label{figure:cleanhcn}
\end{figure}

\begin{figure}
\includegraphics[width=7.5cm]{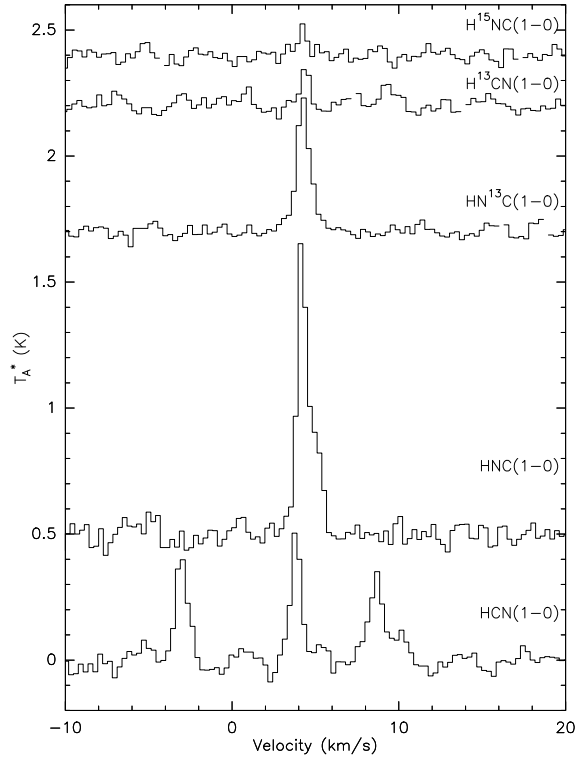}
\caption{Observations used for the column density estimates.  The
position corresponds to the \hnthic\ maximum, which lies at the offset
$(-20^{\prime\prime},+20^{\prime\prime})$ from the dust continuum peak
Cha-MMS1a (marked with a black dot in
Figs.~\ref{figure:hcnmap}-\ref{figure:hn13cmap}).}
\label{figure:5spectra} 
\end{figure}

The excitation temperature is derived by fitting first the optical
thickness, $\tau$, to the observed hyperfine ratios of a ``clean''
\hcn\ spectrum, i.e. a spectrum with a good S/N and hyperfine
intensity ratios in accordance with the LTE assumption.  The $T_{\rm
ex}$ is solved from the antenna equation using observed peak main beam
brightness temperature, $T_{\rm MB}$ and assuming uniform beam
filling. The fit is possible towards the four positions listed in
Table~\ref{table:texfromhcn}. In other positions with strong signal,
like the \hnthic\ maximum $(-20^{\prime\prime},+20^{\prime\prime})$,
or the \hcn\ maximum at $(-20^{\prime\prime},-20^{\prime\prime})$ the
\hcn\ spectra show 'anomalous' hyperfine ratios.  The smallest error
is obtained towards the offset ($40^{\prime\prime},20^{\prime\prime}$)
near the edge of the core with $T_{\rm ex}\approx 5$ K.  This \hcn\
spectrum is shown in Fig.~\ref{figure:cleanhcn}. The other positions
listed in Table~\ref{table:texfromhcn} give similar values of $T_{\rm
ex}$ but the errors are larger.  Also \hcn\ column density estimates
using the values for $\tau$ and $\Delta v$ obtained from hyperfine
fits are given in Table~\ref{table:texfromhcn}.

\begin{table}
\begin{tabular}{rrllll}
$\Delta\alpha$ & $\Delta\delta$ &
$T_{\rm ex}$&$\tau_{\rm main}$&$\Delta v$&$N({\rm HCN})$\\
($^{\prime\prime}$) & ($^{\prime\prime}$) & (K) && (kms$^{-1}$)& (cm$^{-2}$)\\
\hline
 40&  20 &$5.2^{+1.0}_{-0.6}$&$1.1\,\pm0.3$ &$0.49\pm0.30$&$4.7\,10^{12}$\\
-50& -10 &$5.3^{+2.7}_{-1.0}$&$0.9\,\pm 0.4$&$0.58\pm0.07$&$4.6\,10^{12}$\\
 10& -10 &$5.6^{+3.0}_{-1.2}$&$0.7\,\pm 0.3$&$0.56\pm0.08$&$3.8\,10^{12}$\\
 70&  20 &$5.4^{+7.0}_{-1.5}$&$0.5\,\pm 0.3$&$0.49\pm0.04$&$2.0\,10^{12}$\\
\hline
\end{tabular}
\caption{Estimates of the \hcn($J=1-0$) excitation temperatures and 
the \hcn\ column densities towards selected positions using hyperfine 
structure fits. The coordinates are offsets from the position of Cha-MMS1a.}
\label{table:texfromhcn}
\end{table}

The \hcn, \hthicn, \hnthic\ and \hfifnc\ column densities towards the
\hnthic\ maximum, $(-20^{\prime\prime},+20^{\prime\prime})$, have been
derived from the integrated $J=1-0$ line intensities and the $T_{\rm
ex}$ estimate derived from the \hcn\ spectrum shown in
Fig.~\ref{figure:cleanhcn}. For \hcn\ the weakest of the well
separated hyperfine components, $F=0-1$, with the least optical
thickness has been used in this calculation by taking its relative
strength into account.  For \hnc\ we did not attemp to do a similar
estimate because the hyperfine components overlap and are very likely
optically thick.  Overlapping hyperfine components hamper also the
\hnthic\ column density estimate, but their optical thicknesses should
be lower than those of \hnc\ by an order of magnitude or
more. \hfifnc\ has no hyperfine structure due to zero nuclear spin of
$^{15}$N. The results of these calculations are presented in
Table~\ref{table:columndensities}.

\begin{table}
\begin{tabular}{llll}
Molecule&$\int \, T{^*}{_A} dv$&$N_{\rm tot}$\\
 &(${\rm K\,km\,s^{-1}}$)& (${\rm cm}^{-2}$)\\
\hline
\hcn$\,^a$    & 0.37$\,\pm$0.03&$(8.8\pm 1.0)\,10^{12}$\\
\hnc$\,^b$    & 1.17$\,\pm$0.03&                           \\
\hthicn       & 0.19$\,\pm$0.03&$(5.2\,\pm 0.1)\,10^{11}$\\
\hnthic       & 0.52$\,\pm$0.02&$(1.7\,\pm 0.2)\, 10^{12}$\\
\hfifnc       & 0.07$\,\pm$0.01&$(2.3\,\pm 0.5)\, 10^{11}$\\
\hline
\end{tabular}
\caption{Column densities towards the \hnthic\ maximum derived using 
the LTE method. The assumed value for $T_{ex}$ is $5.2^{+1.0}_{-0.6}$ K.
~~$^a$ The intensity value is for the weakest hyperfine component $F=0-1$ which
comprises one ninth of the total intensity of the $J=1-0$ line.
~~$^b$ A column density estimate is not given because of the obvious 
large optical thickness.}  

\label{table:columndensities}
\end{table}

The following column density ratios can be derived from 
Table~\ref{table:columndensities}:

\begin{eqnarray}
\frac{N[\hnthic]}{N[\hthicn]} &=&  3.3\, \pm0.4   \nonumber \\
\frac{N[\hnthic]}{N[\hfifnc]} &=&  7.4\, \pm 1.8  \nonumber \\
\frac{N[\hcn]}{N[\hthicn]} &=&  17 \, \pm 2 \; . \nonumber 
\end{eqnarray}

The \hcn/\hthicn\ column density ratio, 17, is much lower than the
isotopic $^{12}$C/$^{13}$C ratios usually observed in nearby molecular
clouds or local diffuse ISM ($\sim 60$, e.g. \cite{lucas1998}; the
terrestrial value is 89), and is likely to be affected by absorption
or scattering in the less dense envelope of the core. Radiative
transfer effects are even more obvious in the case of the
\hnc/\hnthic\ intensity ratio, 1.7. We do not use therefore the values
derived from the \hcn\ and \hnc\ spectra towards the core centre. The
\hcn\ and \hnc\ spectra are discussed at the end of Sect.~4.
 
\vspace{2mm}

\begin{figure}
\resizebox{\hsize}{!}{\includegraphics[bb=90 360 550 720]{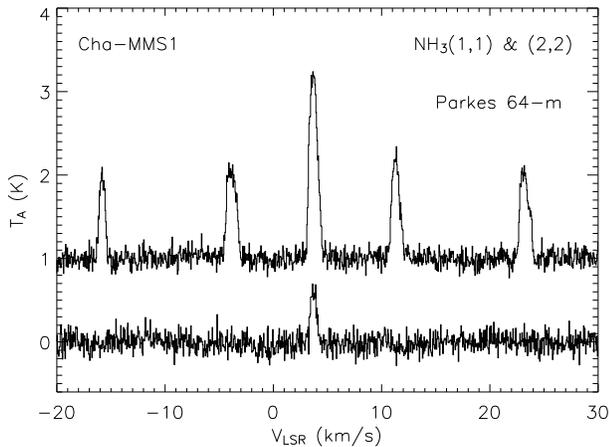}}
\caption{\ammo $(J,K) = (1,1)$ and $(2,2)$ inversion spectra towards
Cha-MMS1a.}
\label{figure:ammonia} 
\end{figure}

The ammonia column density, $N(\ammo)$, and the gas kinetic
temperature, $T_{\rm kin}$, were derived using hyperfine fits to the
\ammo\ inversion lines arising from the rotational states $(J,K) =
(1,1)$ and $(2,2)$. The spectra are shown in
Fig.~\ref{figure:ammonia}.  The standard method of analysis described
e.g. in \cite{ho1979} gave the following values: $N(\ammo) = (1.4 \pm
0.2) \, 10^{15}$ \percmsq\ and $T_{\rm kin} = 12.1 \pm 0.1$ K.
Here it is assumed that the relative populations of all metastable
levels of both ortho- and para-\ammo\ are determined by thermal equilibrium 
at the $(2,2)/(1,1)$ rotational temperature.
   
\subsection{Monte Carlo modelling}

\begin{figure}
\includegraphics[width=7.5cm]{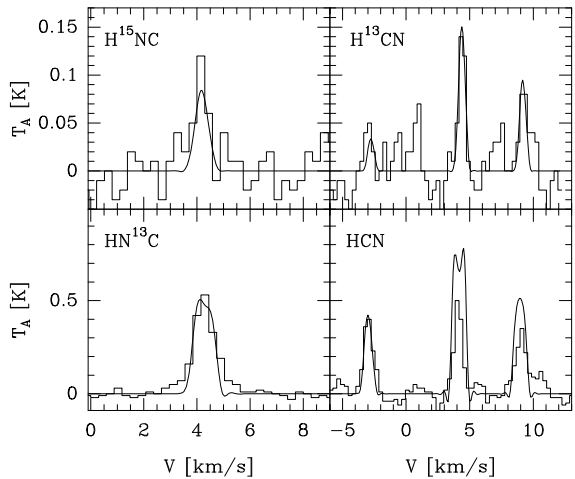}
\caption{Simulated spectra fitted to the observations towards the
\hnthic\ maximum. The assumed kinetic temperature, $T_{\rm
kin}$=12\,K, is based on ammonia observations. The following isotopic
ratios were used in the model calculations: [$^{12}$C]/[$^{13}$C]=20, and
[$^{14}$N]/[$^{15}$N]=280.}
\label{figure:montecarlo}
\end{figure}

\hfifnc, \hnthic, \hcn\ and \hthicn\ spectra were calculated with the
Monte Carlo radiative transfer program of \cite{juvela1997}.
A constant kinetic temperature and a spherically symmetric density 
distribution, n$\sim r^{-1.5}$, were assumed. The selected density power 
law conforms with observational results towards star-forming cores (e.g., 
\cite{tatematsu2004}). In order to avoid divergence in the cloud centre, 
we assume a constant density within a distance corresponding to 6\% of the 
outer radius of the model cloud. Compared with the beam size, the constant 
density region is small and the results are not very sensitive to the selected 
value of its radius. The spectra were convolved with a
gaussian that corresponds to the size of the beam used in the
observations.The simulated spectra were fitted to observations by
adjusting four parameters: the outer radius of the core, the 
scaling of gas density, and the fractional abundances of \hthicn\ and
\hnthic.  The \hcn\ fit was optimized for the weakest hyperfine
component $F=0-1$. The abundances of the other species were determined
by assumed fixed ratios [$^{12}$C]/[$^{13}$C]=20, as suggested by the
LTE column density determination, and [$^{14}$N]/[$^{15}$N]=280
(terrestrial). The resulting fractional \hthicn\ and \hnthic\
abundances do not, however, depend on these assumptions about isotopic
ratios. The calculations were done with $T_{\rm kin}$=12\,K, which was
derived from the ammonia observations, and $T_{\rm kin}$ =20\,K.The
lower temperature resulted in better fit although the difference in
the $\chi^2$ values was no more than 25\%. The results for the 12\,K
are shown in Fig.~\ref{figure:montecarlo}.  The outer radius of the
model is $60^{\prime\prime}$. The volume density increases from
$2.3\,10^4$\,cm$^{-3}$ at the outer edge to $1.4\,10^6$\,cm$^{-3}$ in
the centre, and the molecular hydrogen column density towards the
centre is $N({\rm{H}}_2)=1.9\,10^{22}$\,cm$^{-2}$.

The fractional \hnthic\ and \hthicn\ abundances resulting from the fit
are $2.1\,10^{-10}$ and $5.5\,10^{-11}$, respectively. The
\hnthic/\hthicn\ abundance ratio obtained from the simulations is thus
3.8, i.e. close to the value obtained from the LTE method.

The largest discrepancy between the observed and modelled profiles is
seen in \hcn. It can be explained if the optical depth is 
larger than in this particular model. For example, a
foreground layer of cold gas would readily decrease the intensity of
the \hcn\ lines and would bring the ratios between the hyperfine
components closer to the observed values. The hyperfine ratios are,
however, sensitive to the actual structure of the source
(\cite{gonzales1993}).

The optical thickness of the \hnthic\ main component is $\sim 0.7$.
This moderate value suggests that the method used in the previous
subsection results in a slight underestimatation of column densities.
The \hthicn\ and \hnthic\ column densities (fractional abundances
multiplied by $N({\rm{H}}_2)$) are indeed about two times larger than
those presented in the previous section: $N(\hthicn) = 1.1 \, 10^{12}
\percmsq$, $N(\hnthic) = 4.0 \, 10^{12} \percmsq$. These results suggest, 
however, that the LTE method does not lead to crude underestimates of the 
column densities of rarer isotopologues of HCN and HNC.

\section{Discussion}

\subsection{Fractional abundances}

The present observations and the LTE method give reasonable estimates
for the column densities of \hthicn, \hnthic, \hfifnc\ and \ammo\ near
the centre of Cha-MMS1. Unfortunately the column densities of the
common isotopologues of \hcn\ and \hnc\ cannot be directly determined
because of the large optical thickness of their lines. \hfifnc\
is, however, likely to be useful for this purpose.  According to the
modelling results of \cite{terzieva2000} hardly any $^{15}$N
fractionation occurs in dense interstellar clouds.  The isotopic
$^{14}$N/$^{15}$N ratio lies probably between 240 and 280, i.e. the
values derived for the local ISM (\cite{lucas1998}), and the
terrestrial atmosphere.  Using this range and the \hfifnc\ column
density listed in Table~\ref{table:columndensities} we obtain $N(\hnc)
\sim 6 \, 10^{13} \, \percmsq$. Assuming that no differential $^{13}$C
fractionation occurs, we get from the \hnthic/\hthicn\ ratio derived
above that $N(\hcn) \sim 2 \, 10^{13}$ \percmsq.

The dust continuum emission provides the most reliable estimate for
the \molh\ column density. \cite{reipurth1996} measured towards
Cha-MMS1 a total 1.3 mm flux density of 950 mJy from a nearly circular
region with a diameter of about $50^{\prime\prime}$ (bordered by the lowest
white contour in Fig.~8).  As this size is similar to the SEST
beamsize at 3 mm, the resulting average intensity and average
$N(\molh)$ can be compared with the column densities derived here.
The dust temperature as derived from ISO observations by
\cite{lehtinen2001} is 20 K (Cha-MMS1 = Ced 110 IRS10). Due to
crowding of sources in this region and the limited angular resolution
of ISOPHOT, this value may be affected by emission from other sources
in its vicinity (see Fig. 2 of \cite{lehtinen2001}). Therefore, value
20 K is probably an upper limit.  A lower limit is provided by the gas
kinetic temperature $\sim 12$ K derived above. By using the two temperatures
and Eq. (1) of \cite{motte1998} with $\kappa_{\rm 1.3 mm} = 0.01$
g\percmsq\ appropriate for circumstellar envelopes around Class 0 and
Class I protostars, we get $N(\molh) = 1.9 - 4.0 \,10^{22}\,
{\percmsq}$. The corresponding mass range of the dust core obtained 
from the total flux given by \cite{reipurth1996} is $0.4-0.8\, \Msun$.    

The median of the reasonable H$_2$ column density range, $N(\molh) =
3.0 \,10^{22}\, {\percmsq}$, yields the following values for the
fractional \hcn, \hnc\ and \ammo\ abundances: $\chi(\hcn) \sim 7 \,
10^{-10}$, $\chi(\hnc) \sim 2 \, 10^{-9}$, and $\chi(\ammo) \sim 5 \,
10^{-8}$. The column densities of two other nitrogen bearing molecules
towards Cha-MMS1 were determined by \cite{kontinen2000}: $N(\cyana) =
(4.5 \pm 0.8) \, 10^{14}$ \percmsq, and $N(\diaz) = (1.4 \pm 0.1) \,
10^{13}$ \percmsq, which imply the fractional abundances $\chi(\cyana)
\sim 2 \, 10^{-8}$ and $\chi(\diaz) \sim 5 \, 10^{-10}$.

We find that both \hcn\ and \diaz\ are about 100 times less abundant
than \ammo, whereas \cyana\ is almost equally abundant as \ammo. The
derived \diaz/\ammo\ column density ratio is in accordance with 
previous observational results (e.g., \cite{tafalla2002}; \cite{hotzel2004}). 
The abundance ratios are discussed in Sect. 4.3. 
The \ceio\ column density in this direction is $(1.5 \pm 0.1) \,
10^{15}$ \percmsq\ (\cite{haikala2005}).  The fractional \ceio\
abundance is therefore $5 \, 10^{-8}$ or smaller. This upper limit
falls below the 'standard' value $\sim 2\, 10 ^{-7}$ of
\cite{frerking1982}, and suggests CO depletion.

\subsection{Comparison with previous maps}

\begin{figure}
\resizebox{\hsize}{!}{\includegraphics{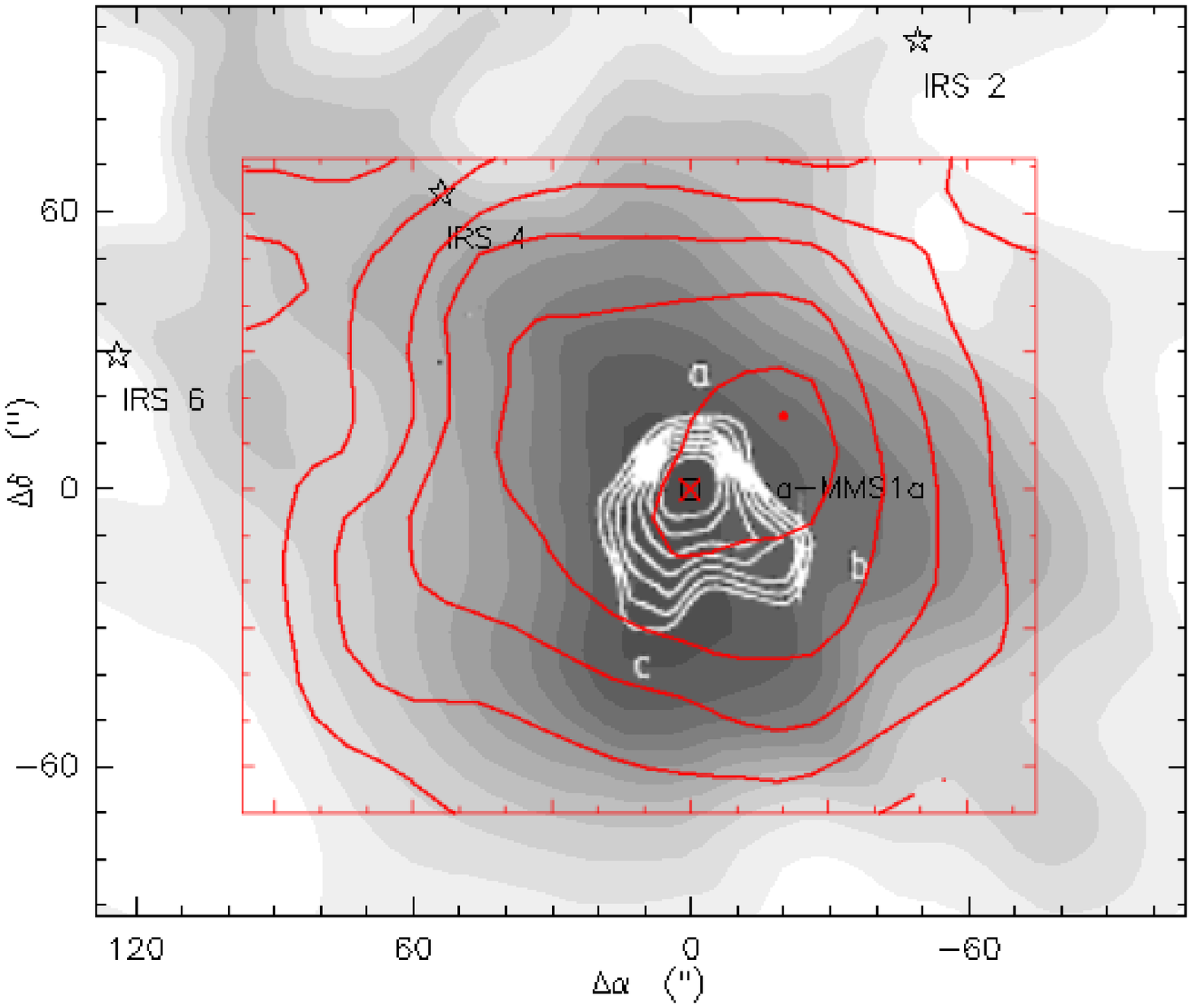}}
\caption{The \hnthic\ map (dark contours) superposed on the \cyana\ map from 
\cite{kontinen2000} (greyscale) and the 1.3 mm dust continuum map of 
\cite{reipurth1996} (white contours).} 
\label{figure:hc3n+dust+hn13c} 
\end{figure}

The superposition of our \hnthic\ map, the \cyana\ map of
\cite{kontinen2000}, and the 1.3 mm dust continuum map of
\cite{reipurth1996} is shown in Fig.~\ref{figure:hc3n+dust+hn13c}.
The two molecular line maps have similar angular resolution (about
$60^{\prime\prime}$) whereas the resolution of the continuum map is
$22^{\prime\prime}$. Both molecules show rather compact, symmetric
distributions around the dust continuum source, \cyana\ agreeing
slightly better with dust than \hnthic. Nevertheless, the dense core
can be clearly localized with the aid of \hnthic\ as well as 
\hcn\ and \hnc. This is in contrast with SO, which shows
two peaks on both sides of the core (\cite{kontinen2000}), and with
C$^{18}$O which has a flat distribution in this region
(\cite{haikala2005}), probably due to depletion. This result
agrees with the results of \cite{hirota2003} and
\cite{nikolic2003}, which suggest that \hcn\ and \hnc\ are less affected
by the accretion onto dust grains than some other commonly observed
molecules.

\subsection{The chemical state of Cha-MMS1 and the HNC/HCN abundance ratio}

CO depletion and a large \ammo\ abundance point towards an advanced
stage of chemical evolution. The \ammo/\diaz\ abundance ratio $\sim
100$ agrees with model predictions for dense cores, but does not put
severe constraints on the evolutionary stage (\cite{aikawa2005}).
According to the observational results of \cite{hotzel2004} this value
should be more appropriate for a star-forming than for a prestellar
core. However, due to slightly different beamsizes the accuracy of the
present determination of the \ammo/\diaz\ ratio is not sufficient for
distinguishing between the two cases.  \cite{kontinen2000} suggested
that the large \cyana\ abundance observed in Cha-MMS1 indicates either
a so called 'late-time' cyanopolyyne peak associated with a large
degree of depletion according to the model of \cite{ruffle1997}, or
carbon chemistry revival by the influence of a newly born, so far
undetected star.  A large ammonia abundance can be understood in
either case.  Ammonia production benefits from the diminishing of CO
and can resist depletion longer (e.g. \cite{nejad1999}). On the other
hand, ammonia can possibly form on grain surfaces and be released into
the gas phase in shocks (\cite{nejad1990}; \cite{aikawa2003}). As
there is no clear evidence for star-cloud interaction in Cha-MMS1, the
alternative that the core - or the portion traced by the present
observations - represents matured pre-stellar chemistry is more
likely.

The behaviour of \hcn\ and \hnc\ under these circumstances is not
clear. The main gas-phase production pathway via the HCNH$^+$,
which is effective especially at later stages, requires \ammo\ and
C$^+$ ions. The latter can be produced also in deep interiors of
molecular clouds via the cosmic ray ionization of C atoms, or in the
reaction between CO and the He$^+$ ion, the latter mechanism becoming
less important with CO depletion (\cite{nejad1999}).  To our knowledge
the production of \hcn\ and \hnc\ on interstellar grains and their
desorption has not been studied.

The models of \cite{aikawa2005} for collapsing prestellar cores
include the gas-phase production of \hcn, \cyana, \ammo, and \diaz.
Although the order of fractional abundances, $\chi(\ammo) > \chi(\hcn)
> \chi(\diaz)$, and the \ammo/\diaz\ abundance ratio derived from the
present observations are well reproduced by this model, especially for
the gravity-to-pressure ratio $\alpha=4$ impying a rapid collapse and
less marked depletion, the predicted \hcn\ column densities relative
to \ammo\ and \diaz\ are an order of magnitude larger than those
observed here.  A likely explanation is that \hcn\ is
not as centrally peaked than \ammo\ and \diaz\ (see Fig.~3 of
\cite{aikawa2005}).  Therefore a large fraction of \hcn\ resides in on
the outskirts of the core, where the density is not sufficient for the
collisional excitation of its rotational levels, and the molecule is
not seen in emission (see also the next section). Another discrepancy
is that the model predicts much lower \cyana\ column densities
relative to ammonia than observed here. The agreement is slightly
better for the model with overwhelmingly large gravitational potential
($\alpha=4$) than for the slowly evolving model near hydrostatic
equilibrium ($\alpha=1.1$) which results in very high degrees of
depletion. Unfortunately the spectral resolution and S/N ratio of the
present data do not allow to study the core kinematics.

As compared to the 'principal' N-bearing molecules \ammo\ and \cyana,
\hcn\ and \hnc\ have clearly smaller abundances, and probably only
marginal importance to the nitrogen chemistry in the
core. Nevertheless, the fact that \hnc/\hcn\ abundance ratio is once
again observed to be larger than unity remains a
problem. \cite{talbi1998} have shown that even though the reaction
between \ammo\ and C$^+$ may produce a significant amount of the
metastable ion ${\rm H_2NC^+}$, which should produce only \hnc\ in
electron recombination, the energy released in the former reaction
leads to the efficient transformation into the linear isomer ${\rm
HCNH}^+$ before the recombination. Likewise, according to
\cite{herbst2000}, the vibrational energies of \hnc\ and \hcn\ formed
in the {\sl neutral} reactions ${\rm C} + {\rm NH_2} \rightarrow \hnc
+ {\rm H}$, and ${\rm N} + {\rm CH_2} \rightarrow \hcn + {\rm H}$ are
sufficient to overcome the isomerization barrier, and roughly equal
amounts of both isomers are produced.
   
The theoretical results presented in \cite{talbi1998} and
\cite{herbst2000}, and more recently by \cite{hickman2005} and
\cite{ishii2006}, show quite convincingly that the \hnc/\hcn\
abundance ratio resulting from the chemical reactions known to be
effective in dark clouds {\sl cannot} differ significantly from unity.
There are two possible explanations for the discrepancy between the
observations and theoretical predictions. Either so far unknown
processes favour \hnc\ in the cost of \hcn, or then the derived
\hnthic/\hthicn\ column density ratio is affected by different degrees
of $^{13}$C fractionation in \hnc\ and \hcn. In the latter case
$^{13}$C would about three times more enhanced in \hnc\ than in \hcn\
in Cha-MMS1.
   
The \hnthic/\hfifnc\ column density ratio 7.4 together with the above
quoted range of $^{14}$N/$^{15}$N ratios imply that the \hnc/\hnthic\
ratio in Cha-MMS1 is in the range $30-40$. As the typical
$^{12}$C/$^{13}$C isotopic ratio observed in nearby clouds is 60
(\cite{lucas1998}) the result suggests considerable $^{13}$C
enhancement in \hnc. An \hcfifn\ column density determination would
readily show, according to the results of \cite{terzieva2000}, whether
also \hcn\ has a similar degree of $^{13}$C fractionation.

In the only survey so far including all these isotopologues,
\cite{irvine1984} found very similar \hnthic/\hfifnc\ and
\hthicn/\hcfifn\ intensity ratios ($\sim 1.5$) towards TMC-1. This
observation suggests similar $^{13}$C fractionation in \hcn\ and
\hnc. On the other hand, this result might not be universally valid.
\cite{ikeda2002} determined substantially different \hthicn/\hcfifn\
column density ratios in the starless dark cloud cores L1521E, TMC-1,
and L1498, listed in order of increasing \hthicn/\hcfifn\ ratio (2.6,
6.7, and 11).  The \hnthic/\hthicn\ column density ratios in these
sources increase in the same order (0.54, 1.7, and 4.5;
\cite{hirota1998}).  The values obtained towards TMC-1 are consistent
with those of \cite{irvine1984}, but the results towards L1521E and
L1498 suggest (if the possibility of $^{15}$N fractionation is
neglected) that the $^{13}$C fractionation in \hcn\ changes from
source to source, and perhaps the variation in the case of \hnc\ is
even larger.  L1521E is suggested to represent youthful chemistry with
little molecular depletion (\cite{hirota2002}; \cite{tafalla2004})
whereas the position observed in L1498 is known to be heavily depleted
(\cite{willacy1998}).  Thus, it appears that the freeze-out of
molecules favour the $^{13}$C substituted isotopologues of \hcn\ and
\hnc.  It seems warranted to further examine the possibility of their
differential $^{13}$C fractionation, although no theoretical arguments
are in favour of different gas-phase production and destruction rates
of \hnthic\ and \hthicn, or their different freezing rates. As no
empirical data on this matter is currently available, the best way to
proceed is probably to perform similar observations as those done by
\cite{irvine1984} towards a sample of cores with different degree of
depletion.

\subsection{Scattering envelope}

The low \hcn/\hthicn\ and \hnc/\hnthic\ intensity ratios and the line
shapes in the \hcn\ and \hnc\ spectra observed towards the centre of
the core require some attention.  The intensity ratios suggest very
large optical thicknesses for the \hcn\ and \hnc\
lines, and consequently, deep self-absorption features and 'anomalous'
hyperfine component ratios are to be expected.  A clear absorption
feature can be seen in the {\sl simulated} \hcn\ spectrum shown in
Fig.~\ref{figure:montecarlo}, although a very low $^{12}$C/$^{13}$C
isotopic ratio is used in order to avoid inverted hyperfine ratios,
which would disagree with the observations.  The {\sl observed} \hcn\
and \hnc\ spectra are, however, free from absorption dips, and they
look more like weakened 'normal' lines.

This weakening can probably be attributed to less dense gas
surrounding the core. The \hcn\ and \hnc\ molecules in the envelope,
where the density lies below the critical density of their $J=1-0$
lines, function as scatterers of line emission arising from the core,
i.e. they re-emit absorbed photons (into random directions) before
colliding with other particles.  The effect is less marked for $^{13}$C
isotopologues, and the resulting \hcn/\hthicn\ and \hnc/\hnthic\
intensity ratios decrease, without \hcn\ and \hnc\ loosing their
characteristic line patterns. This model was used by
\cite{cernicharo1984} to explain \hcn\ hyperfine anomalies,
particularly the intesity of the intrinsically weakest component
$F=0-1$ observed towards dark clouds. The \hcn\ spectrum shown in
Fig.~\ref{figure:5spectra} conforms with this explanation: the $F=0-1$
component on the left is stronger than $F=1-1$ component on the right,
although the optically thin LTE ratio is $1:3$.

The scattering of \hcn\ emission from the core has probably lead to
underestimation of the excitation temperature and column densities
derived using the LTE method in Sect.~3.2. For example, an increase of
1 K in $T_{\rm ex}$, would mean 30 \% larger column densities, and 3 K
would correspond to a 100\% increase in the column density.
Quantitative estimates of the scattering effects are, however, beyond
the scope of this paper, and here we merely note this additional
source of error.

\section{Summary}

A mapping in the $J=1-0$ rotational lines of \hcn, \hnc\ and \hnthic\
shows a compact structure in the direction of the protostellar core
Cha-MMS1. The \hnthic\ map agrees best with the previous 1.3 mm dust
continuum map of \cite{reipurth1996}, and the \cyana\ map of
\cite{kontinen2000}.

The column density determinations of \hthicn, \hnthic, \hfifnc\ and
\ammo, and the results of the quoted two studies allow us to estimate
that the fractional \hcn\ and \hnc\ abundances are $7 \, 10^{-10}$ and
$2 \, 10^{-9}$, respectively. These are slightly larger than that of
\diaz, but clearly smaller than the \ammo\ and \cyana\ abundances ($3
\, 10^{-8}$ and $1 \, 10^{-8}$, respectively). In accordance with
recent results in other dense cores we find that \ammo\ is about 100 times
more abundant than \diaz. The kinetic temperature derived from the
\ammo\ lines is about 12 K. The derived fractional abundances conform
with the previous suggestion of \cite{kontinen2000}, that Cha-MMS1
represents an advanced chemical state, although it has a large \cyana\
abundance, which usually is thought to indicate early times. This
situation is likely to have developed under the circumstances of CO
depletion (\cite{ruffle1997}). 

The following column density ratios are found: $\hnthic/\hthicn \,
\sim 3-4$ and $\hnthic/\hfifnc \, \sim 7$.  Using the latter ratio,
and the assumption based on the modelling results of
\cite{terzieva2000}, we estimate that the [{\hnc}]/[{\hnthic}] ratio
is between 30 and 40. As there is no \hcfifn\ column density
determination available towards this source, the possibility that
differential $^{13}$C fractionation causes the observed relatively
high [{\hnthic}]/[{\hthicn}] ratio cannot be entirely ruled out. 

The \hcn\ and \hnc\ lines are not much brighter than those of the
$^{13}$C substituted species, but still show no absorption dips. These
lines are probably weakened by scattering by \hcn\ and \hnc\ molecules
residing in less dense gas around the core. As suggested by
\cite{cernicharo1984}, the same process can cause the so called \hcn\
hyperfine anomalies in dark clouds, and is probably also responsible
for the 'too' bright $F=0-1$ component of HCN($J=1-0$) towards the
centre of Cha-MMS1. The evidence for scattering and the
relatively low \hcn\ and \hnc\ column densities derived towards the
core suggest that the \hcn\ and \hnc\ distributions are less centrally
peaked than those of \ammo\ and \diaz, which would be in agreement
with the model of \cite{aikawa2005} for a rapidly collapsing core.

\vspace{3mm}

We thank Wolf Geppert for useful discussions and the anonymous referee 
for valuable comments on the manuscript. 
This project was supported by the Academy of Finland, grant
Nos. 173727 and 174854. P.P.T. thanks the ERASMUS programme of the 
European Union, Olga Koningfonds, Utrecht, and Kapteyn Fonds, Groningen.

\end{document}